\documentclass[aip,jmp,showkeys,amsmath,amssymb,superscriptaddress]{revtex4}

\usepackage{natbib,xcolor}

\begin{document}

\title{Entropy and complexity analysis of hydrogenic Rydberg atoms}

\author{S. L\'opez-Rosa}
\email{slopezrosa@us.es}
\affiliation{Instituto Carlos I de F\'isica Te\'orica y Computacional, Universidad de Granada, 18071-Granada, Spain}
\affiliation{Departamento de F\'isica Aplicada II, Universidad de Sevilla, 41012-Sevilla, Spain}

\author{I.V. Toranzo}
\email{irevato@correo.ugr.es}
\affiliation{Instituto Carlos I de F\'isica Te\'orica y Computacional, Universidad de Granada, 18071-Granada, Spain}
\affiliation{Departamento de F\'isica At\'omica, Molecular y Nuclear, Universidad de Granada, 18071-Granada, Spain}

\author{P. S\'anchez-Moreno}
\email{pablos@ugr.es}
\affiliation{Instituto Carlos I de F\'isica Te\'orica y Computacional, Universidad de Granada, 18071-Granada, Spain}
\affiliation{Departamento de Matem\'atica Aplicada, Universidad de Granada, 18071-Granada, Spain}

\author{J.S.Dehesa}
\email{dehesa@ugr.es}
\affiliation{Instituto Carlos I de F\'isica Te\'orica y Computacional, Universidad de Granada, 18071-Granada, Spain}
\affiliation{Departamento de F\'isica At\'omica, Molecular y Nuclear, Universidad de Granada, 18071-Granada, Spain}

\date{\today}

\begin{abstract}
The internal disorder of hydrogenic Rydberg atoms as contained in their position and momentum probability densities is examined by means of the following information-theoretic spreading quantities: the radial and logarithmic expectation values, the Shannon entropy and the Fisher information. As well, the complexity measures of  Cr\' amer-Rao, Fisher-Shannon and LMC types are investigated in both reciprocal spaces. The leading term of these quantities is rigorously calculated by use of the asymptotic properties of the concomitant entropic functionals of the Laguerre and Gegenbauer orthogonal polynomials which control the wavefunctions of the Rydberg states in both position and momentum spaces. The associated generalized Heisenberg-like, logarithmic and entropic uncertainty relations are also given. Finally, application to linear ($l=0$), circular ($l=n-1$) and quasicircular ($l=n-2$) states is explicitly done.
\end{abstract}


\maketitle

\section{Introduction}

Although named for the 19th century Swedish spectroscopist Johannes Rydberg, the Rydberg atoms (i.e. atoms in which an electron has been excited to an unusually high energy level, so with a very high principal quantum number $n$) were first detected in interstellar space in 1965; a few years later, with the advent of precisely tunable dye lasers it was possible to make Rydberg atoms routinely, pushing electrons out to larger and larger orbits. Since then, they have been increasingly studied mostly for two reasons. First, they are stepping stones from the quantum to classical worlds since they lie at the border between bound states and the continuum; so that, any process which can result in excited bound states or ions and free electrons usually leads to the production of Rydberg states. The Rydberg states of atoms and molecules \cite{stebbings_83, gallagher_94, gallagher_06, lundee:aamop05, wilson_03, connerade_98} constitute a fertile laboratory where to investigate the order-to-chaos transitions through the applications of electric fields \cite{shiell_03}. Second, the extraordinary properties  of these atoms (gigantic size, extremely long lifetimes and strong polarization by relatively weak electric fields,...) allow experiments to be done which would be difficult or impossible with normal atoms. This has been recently illustrated in the applicability of the Rydberg atoms for quantum information processing \cite{shiell_03, saffman:rmp10}.

For Rydberg atoms with more than one electron, the outer electron is excited into such a high-lying energy level; so that it moves in a spatially extended orbital far outside the charge distribution of the other electrons. The Rydberg electron feels an atomic core with an effective charge $+e$, resulting of the nucleus and all the inner electrons, so that the atom behaves in many respects like a hydrogen atom as long as the Rydberg electron does not approach the core too closely. The latter in turn depends on the angular momentum of the Rydberg electron. For the maximum angular momentum quantum number $l=n-1$, the classical Rydberg orbit is circular and the penetration depth of the electron into the core is minimum, while for small values $l<<n$ the orbit becomes elliptical with large eccentricity in which the potential felt by the electron deviates from the Coulombian one.

In this paper we consider both Rydberg states of hydrogen atoms ($Z=1$) and Rydberg states of ions and molecules, where one electron is highly excited, but still weakly bound to positively charged ion which represents the remainder of the system. The Rydberg electron is primarily sensitive to the total charge of the core formed by the nucleus and all the inner electrons, which binds it in its orbit. Associated to each large principal quantum number $n$, there exists a manifold of such Rydberg states, analogous to elliptical orbits sharing the length of their major axis but ranging in shape from linear ($l=0$) to circular ($l=n-1$) and quasicircular ($l=n-2$).

The goal of this paper is to investigate the internal disorder of the Rydberg systems by means of the analytical information theory. This includes the quantification of the multiple facets of the spatial and momentum extension, as given by the three-dimensional geometries of the charge and momentum probability distributions of the corresponding Rydberg states,  by means of the following spreading measures: power moments, variances, logarithmic expectation values, Shannon entropies \cite{shannon:bst48} and Fisher informations \cite{frieden_04, fisher:pcps25} and some intrinsic or density-dependent complexity measures of Cr\' amer-Rao \cite{dembo:ieetit91, dehesa:jcam06, ANT}, Fisher-Shannon \cite{ANG, ROM} and LMC (L\'opez Ruiz-Mancini-Calvet) \cite{lopezruiz_pla95,catalan:pre02} types.

The electronic probability densities of the Rydberg atoms are given \cite{bethe_57, hey:ajp93, hey:ajp93b, fock:zp35, lombardi:pra80} by the expressions 
\begin{equation}\label{eq:rho}
\rho(\vec{r})= \frac{4 Z^3}{n^4} \frac{\omega_{2l+1}(\tilde{r})}{\tilde{r}} \left[{\tilde{\cal{L}}}_{n-l-1}^{2l+1}(\tilde{r}) \right]^2 \left| Y_{l,m}(\theta,\varphi) \right|^2,
\end{equation}
(with $\tilde{r}=\frac{2Z}{n} r$, $l=0,1,\dots,n-1$, and $m=-l,-l+1,\dots,+l$ and  $\omega_{2l+1}(x)=x^{2l+1} e^{-x}$) for the electronic orbitals in configuration or position space, and
\begin{equation}\label{eq:gamma}
\gamma(\vec{p})= 2^{2l+4} \left( \frac{n}{Z}\right)^3 \frac{(n \tilde{p})^{2l}}{(1+n^2 {\tilde{p}}^2)^{2l+4}} \left[{\tilde{\cal{C}}}^{l+1}_{n-l-1} \left(\frac{1-n^2 {\tilde{p}}^2}{1+n^2 {\tilde{p}}^2} \right) \right]^2  \left| Y_{l,m}(\theta,\varphi) \right|^2,
\end{equation}
(with $\tilde{p}=\frac{p}{Z})$ for the electronic orbitals in reciprocal or momentum space. The symbol $ Y_{l,m}(\theta,\varphi)$ denotes the ordinary three-dimensional spherical harmonics. Moreover the symbols ${\tilde{\cal{L}}}_{m}^{\alpha}$ and ${\tilde{\cal{C}}}_{m}^{\alpha}$ denote the Laguerre and Gegenbauer polynomials, orthonormal with respect to the weight functions $\omega_{\alpha}(x)=x^\alpha e^{-x}$ and $\omega^{*}_{\alpha}(x)=(1-x^2)^{\alpha-\frac{1}{2}}$, respectively.

The physical and chemical properties of the Rydberg atoms are completely determined by the spread of the position and momentum densities (\ref{eq:rho}) and (\ref{eq:gamma}) all over the space. The quantification of the multiple facets of this spreading is presently done not only by means of the radial expectation values in position, $\langle r^k \rangle$,  and momentum, $\langle p^k \rangle$, spaces but also by some information-theoretic measures of local (Fisher's informations, $I\left[ \rho \right]$ and $I\left[ \gamma \right]$) and global (Shannon's entropies, $S\left[ \rho \right]$ and $S\left[ \gamma \right]$) character. These spreading measures are closely related to various fundamental and/or experimentally measurable quantities such as the diamagnetic susceptibility ($\langle r^2 \rangle$), the potential energy ($\langle r^{-1} \rangle$), the kinetic energy ($\langle p^2 \rangle$), etc. Moreover, they allow to determine various uncertainty measures and to bound the macroscopic quantities which are described by power density functionals of the system. 

Let us highlight the role of the Shannon and Fisher uncertainty measures. While the former quantifies the global extent of the density (since it is a logarithmic functional of it), the Fisher information is a local uncertainty measure which quantifies the gradient content (so, the oscillatory character) of  the Rydberg wavefuncion because it is given by a gradient functional of the corresponding density. Moreover, both measures fulfil a number of interesting properties: (a) they are the basic variables of two extremization procedures (the maximum entropy method and the principle of extreme physical information \cite{frieden_04}), (b) they have been used to identify the most distinctive non-linear phenomena (avoided crossings) encountered in atomic and molecular spectra under external fields \cite{gonzalezferez:prl03}, and (c) they satisfy the uncertainty relations \cite{bialynickibirula:cmp75, sanchezmoreno:jpa11}:
\begin{equation}\label{eq:S_rhoS_gamma}
S \left[\rho \right] + S \left[\gamma \right] \geq 3 \left( 1+ \ln \pi \right),
\end{equation}
\begin{equation}\label{eq:I_rhoI_gamma}
I \left[\rho \right] \times I \left[\gamma \right] \geq 36,
\end{equation}
which generalize the well-known Heisenberg-like relation $\displaystyle{\left\langle r^2 \right\rangle \left\langle p^2 \right\rangle \geq \frac{9}{4}}$ and its extension based upon moments of arbitrary order \cite{zozor:pra11,angulo:pra94}. For further results, see the reviews \cite{thakkar:acp04,dehesa_10,gadre_03}.

More recently, some density-dependent complexity measures, such as the Cr\'amer-Rao, Fisher-Shannon, LMC  and LMC-R\'enyi generalized complexities have been introduced in a quantum-physical context \cite{romera_irp09,lopezruiz_jmp09,nagy_pla09,sanudo_pla08}; see also the recent reviews \cite{dehesa_10,AN1}.
Contrary to other complexity notions (computational complexity, algorithmic complexity,...) which one may eventually think of, the density-dependent complexities are intrinsic properties of the quantum system. Moreover, the effective complexity of the system in a given quantum state is closely related to the main features of the associated quantum-mechanical probability density $\rho(\vec{r})$ (irregularities, extent, fluctuations, smoothing,...). 

Each of these measures of complexity grasps the combined balance of two different spreading facets of the probability density. The Cr\' amer-Rao complexity quantifies the gradient content of $\rho(\vec{r})$ jointly with the probability spreading around the centroid. The Fisher-Shannon complexity measures the gradient content of $\rho(\vec{r})$ together with its total extent in the position space. The LMC complexity measures the combined balance of the average height of $\rho(\vec{r})$ (as given by the second-order entropic moment $W_{2}[\rho]$, also called disequilibrium $D[\rho]$), and its total extent (as given by the Shannon entropic power $N[\rho] = e^{S[\rho]})$. Moreover, it is interesting to notice that these three complexity measures are (a) dimensionless, (b) bounded from below by unity (when $\rho$ is a continuous density in $R$ in the Cr\' amer-Rao and Fisher-Shannon cases, and for any $\rho$ in the LMC case), and (c) minimum for the two extreme (or least complex) distributions which correspond to perfect order (i.e. the extremely localized Dirac delta distribution) and maximum disorder (associated to a highly flat distribution). Finally, they fulfil invariance properties under replication, translation and scaling transformation \cite{YA1, YA2}.

The structure of this paper is the following. Firstly, in Section \ref{sec:radial_exp}, we show the known position and momentum radial expectation values of the Rydberg atoms recently obtained by use of some modern ideas and techniques relative to the asymptotics of the Laguerre and Gegenbauer polynomials of varying kind (i.e., when the parameter of the polynomials are dependent on their degree). Moreover, for the sake of completeness, we also consider the logarithmic expectation values, and the associated uncertainty products. Then, we analyze the Shannon entropy and the Fisher information of the Rydberg atoms for both position and momentum spaces in  Sections \ref{sec:shannon} and \ref{sec:fisher}, respectively. In Section V we calculate and discuss  the
complexity measures of the Rydberg atoms mentioned above. In Section \ref{sec:circular} we apply the previous general results to some specific states of the Rydberg atoms of circular $(l = n-1)$ and quasi-circular $(l = n-2)$ types. Finally some conclusions and open problems are pointed out. 

\section{Position and momentum expectation values: BASICS}
\label{sec:radial_exp}

In this Section we show the known values of the position and momentum radial expectation values ($\left\langle r^\alpha \right\rangle$ and $\left\langle p^\alpha \right\rangle$, $\alpha \in  \mathbb{R}$, respectively) of the Rydberg atoms, pointing out some open problems. As well, we find the logarithmic expectation values of these systems, denoted by $\left\langle \ln r \right\rangle$ and $\left\langle \ln p \right\rangle$ respectively. Finally, we show that the Heisenberg-like product ($\left\langle r^{\alpha} \right\rangle\cdot\left\langle r^{\beta} \right\rangle$) and the logarithmic sum ($\left\langle \ln r \right\rangle + \left\langle \ln p \right\rangle $) fulfil not only the Heisenberg-like and Beckner logarithmic \cite{beckner:ams75} uncertainty relations, valid for general quantum systems, but also the corresponding more stringent uncertainty relations valid for quantum systems with central potentials recently obtained by Zozor et al \cite{zozor:pra11} and Rudnicki et al \cite{rudnicki_12}, respectively. \\
The radial expectation values in position space, $\left\langle r^\alpha \right\rangle$ with $r= \| \vec{r} \|$, of the Rydberg atom in the state $(n,l,m)$ are given by 
\begin{eqnarray}
\left\langle r^\alpha \right\rangle &=& \int r^\alpha \rho(\vec{r}) d \vec{r} \\
														&=& \frac{1}{2n} \left( \frac{n}{2Z} \right)^{\alpha} \int_{0}^{\infty} \omega_{\nu}(\tilde{r}) \left[ {\tilde{\cal{L}}}_{k}^{\nu}(\tilde{r}) \right]^2 {\tilde{r}}^{\alpha+1} d\tilde{r}
\end{eqnarray}
with $k = n-l-1$ and $\nu  =  2l+1$, where we have taken into account the expression (\ref{eq:rho}) for the probability density $\rho(\vec{r})$ of the state with quantum numbers $(n,l,m)$. Notice that this integral converges for all values of $\alpha > -2l-3$, and the Laguerre polynomial is of varying type (i.e., the parameter ν depends on the degree since $l \in  \left\lbrace 0, 1,….., n-1 \right\rbrace$). Moreover, remark that $k\rightarrow \infty$ if and only if $n \rightarrow \infty$. Recently, using the Buyarov et al's theorem \cite{buyarov:jat99}) which gives the weak-* asymptotics of the Laguerre polynomials, Aptekarev et al \cite{aptekarev:jpa10} have shown that

\begin{equation}
\label{eq:r_alpha_res}
\left\langle r^\alpha \right\rangle \underset{n\rightarrow \infty}{\backsimeq} \left\{ \begin{array}{lcc}
           																		  \left( \frac{n^2}{Z} \right)^{\alpha} \frac{2^{\alpha+1} \Gamma \left(\alpha + \frac{3}{2} \right)}{\sqrt{\pi} \Gamma \left(\alpha+2 \right)} &   ,  & \alpha > -\frac{3}{2} \\
            												 \\\frac{Z^{-\alpha}}{n^3} \frac{ \Gamma \left(2l+3+\alpha\right)}{\Gamma \left(2l-\alpha \right)} \frac{2^{-(3\alpha+5)} \Gamma \left(-\alpha - \frac{3}{2} \right)}{\sqrt{\pi} \Gamma \left(-\alpha-1 \right)} &  , & -2l-3 < \alpha < -\frac{3}{2}  
          																					   \end{array}
 																								  \right.
\end{equation}
what completes the full range of admissible powers of $\left\langle r^\alpha \right\rangle$, except when $\alpha=-\frac{3}{2}$ (which remains to be an open problem). For checking and completeness, let us give the first few expectation values 
\[
\left\langle r^0 \right\rangle= 1, \quad \left\langle r \right\rangle \backsimeq \frac{3n^2}{2Z}, \quad \left\langle r^2 \right\rangle \backsimeq \frac{5}{2} \left( \frac{n^2}{Z} \right)^2,
\]
and
\begin{align*}
&\left\langle r^{-1} \right\rangle= \frac{Z}{n^2} \\
& \left\langle r^{-2} \right\rangle= \frac{Z^2}{n^3 \left(l+\frac{1}{2} \right)}, \qquad \left\langle r^{-3} \right\rangle=  \frac{Z^3}{n^3 l \left(l+1 \right)\left(l+\frac{1}{2} \right)},\nonumber \\
& \left\langle r^{-4} \right\rangle \backsimeq \frac{3 Z^4}{2 n^3 l \left(l+\frac{3}{2} \right)\left(l+1 \right)\left(l+\frac{1}{2} \right)\left(l-\frac{1}{2} \right) },
\end{align*}
with integer powers $\alpha$ for all Rydberg states $(n \rightarrow \infty)$. These first few expectation values can be tested with the exact values given in Ref. \cite{drachman:pra82} and \cite{gallagher_94}, respectively. Let us also highlight that Eq. (\ref{eq:r_alpha_res}), which give the exact leading term of the radial expectation values of the Rydberg states with $(l,m)$ fixed, improve the corresponding approximate values obtained by Shiell \cite{shiell_03} from simple geometrical considerations and extend and rigorously prove the semiclassical values obtained by Heim \cite{heim:jpb94}.

Let us now calculate the radial expectation values of Rydberg atoms in momentum space given by
\begin{align*}
\left\langle p^\alpha \right\rangle & = \int p^\alpha \gamma(\vec{p}) d \vec{p}; \qquad \quad \alpha \in  \mathbb{R} \nonumber \\
&= \left( \frac{Z}{n} \right)^{\alpha} \int_{-1}^{+1} \omega_{\nu'}^{*}(t) \left[ {\tilde{\cal{C}}}_{k}^{(\nu')}(t) \right]^2 \left(  1-t \right)^{\frac{\alpha}{2}} \left(  1+t \right)^{1-\frac{\alpha}{2}}dt,
\end{align*}
where $\omega_{\nu'}^{*}(t)$ is the weight function of the Gegenbauer polynomials ${\tilde{\cal{C}}}_{k}^{(\nu')} (t)$, $k=n-l-1$ and $\nu'=l+1$. We first note that this integral converges only when $-2 \nu'-1 < \alpha < 2\nu'+3$, i.e., when $-2l-3 < \alpha < 2l+5$. 
In 2010 Aptekarev et al \cite{aptekarev:jpa10} have shown by means of the weak-* asymptotics of Gegenbauer polynomials \cite{buyarov:jat99} that the momentum expectation values $\left\langle p^\alpha \right\rangle$, $-1<\alpha<3$ and $\alpha \neq 1$, of the Rydberg states $(n,l,m)$ with $l$ uniformly bounded, have the expression
\begin{equation*}
\left\langle p^\alpha \right\rangle \backsimeq \frac{1}{\pi} \left( \frac{Z}{n} \right)^{\alpha}  \int_{-1}^{+1} \left( \frac{1-t}{1+t}\right)^{\frac{\alpha-1}{2}} dt,
\end{equation*}
which gives the following momentum expectation values for the Rydberg atoms
\begin{align}\label{eq:p_alpha_n-infty2}
&\left\langle p^{\alpha} \right\rangle \backsimeq \left( \frac{Z}{n} \right)^{\alpha}  \frac{\alpha-1}{ \sin \left(\frac{\pi}{2} (\alpha-1) \right)}=\left( \frac{Z}{n} \right)^{\alpha} \frac{2}{\pi} \Gamma \left( \frac{\alpha+1}{2} \right)\Gamma \left( \frac{3-\alpha}{2} \right); \quad -1< \alpha < 3, \quad \alpha \neq 1, \\
& \left\langle p \right\rangle  \backsimeq \frac{2 Z}{\pi n}, \nonumber
\end{align} 
which provides, in particular, the exact values for the normalization $ \left( \left\langle p^0\right\rangle = 1\right)$ and the kinetic energy $ \left( \left\langle p^2 \right\rangle = Z^2/n^2 \right)$ of the system.  

Let us note that, as it happened in the position case, here we are able to obtain the leading term of the momentum expectation values $\left\langle p^\alpha \right\rangle$ of the Rydberg atoms only in a subrange of all of possible values of $\alpha$ $(-2l-3<\alpha<2l+5)$; namely, in the subrange $\alpha \in (-1,3)$. Again, this is related to the use of the weak-* asymptotics for the involved polynomials instead of the strong asymptotics in the spirit of \cite{aptekarev:rassm95, dehesa:jmp98}, but the latter lie out of the scope of this work. Unfortunately, here we do not have a recursion relation of the position type  for the momentum expectation values to overcome this problem and extend Eq. (\ref{eq:p_alpha_n-infty2}) beyond the interval $(-1, 3)$. Let us only mention here that Delburgo and Elliot have obtained the expectation value $\left\langle p^{-1} \right\rangle$ \cite{delbourgo:jmp09}; the determination of the remaining possible expectation values is an open problem.

For $s=0$ ($l$ finite value), the generalized Heisenberg-like product becomes
\begin{equation*}
\left\langle r^\alpha \right\rangle^\frac{2}{\alpha}  \left\langle p^\beta \right\rangle^\frac{2}{\beta} =
\left(
\frac{2^{\alpha+1}\Gamma\left(\alpha+\frac{3}{2} \right)}{\sqrt{\pi}\Gamma(\alpha+2)}
\right)^\frac{2}{\alpha}
\left(
\frac{2}{\pi}  \Gamma\left(\frac{1+\beta}{2} \right) \Gamma\left(\frac{3-\beta}{2} \right)
\right)^\frac{2}{\beta}
n^2+ o (n^{2}),
\end{equation*}
valid for $\alpha>-\frac{3}{2}$ and $-1<\beta<3$. Let us point out that this uncertainty product fulfils the generalized Heisenberg-like uncertainty relation recently found by Zozor et al \cite{zozor:pra11}.

The logarithmic expectation values $\left\langle \ln r \right\rangle$ and $\left\langle \ln p \right\rangle$ of Rydberg atoms, which has been shown to play a very useful role in the model-independent description of atomic, molecular and nuclear charge distribution \cite{friedrich:npa72,andrae:pr00}, can be much easier calculated by taking into account the numerous known algebraic properties of the involved Laguerre and Gegenbauer polynomials. Their values, following the lines of \cite{dehesa:ijqc10}, turn out to be given by
\begin{equation*}
\left\langle \ln r \right\rangle =  2 \ln n+1-\ln 2-\ln Z+O(n^{-2}),
\end{equation*}
in position space, and
\begin{equation*}
\left\langle \ln p \right\rangle =  -\ln n-1+\frac{l+\frac{1}{2}}{n}+\ln Z+O(n^{-2}),
\end{equation*}
so that the net uncertainty logarithmic sum has the value
\begin{equation*}
\left\langle \ln r \right\rangle+\left\langle \ln p \right\rangle = \ln n - \ln 2 + \frac{l+\frac{1}{2}}{n}+O(n^{-2}),
\end{equation*}
which certainly fulfills not only the Beckner uncertainty relation \cite{beckner:ams75} valid for general systems, but also the improved relation $\langle \ln r\rangle + \langle \ln p \rangle \ge \psi \left(\frac{2l+3}{4}\right)+\ln 2$, $l=0,1,2,\ldots$, valid for any central potential, recently found by Rudnicki et al \cite{rudnicki_12}, where $\psi(x)$ is the psi or digamma function.

\section{Position and momentum Shannon entropies}
\label{sec:shannon}

The extent to which a probability density is spread all over the space, is best determined by its Shannon entropy. In this Section, we calculate the leading term $(n \rightarrow \infty)$ of the Shannon entropies of the electronic densities $\rho(\vec{r})$ and $\gamma(\vec{p})$ of Rydberg atoms in position and momentum spaces, given by Eqs. (\ref{eq:rho}) and (\ref{eq:gamma}), explicitly in terms of the three quantum numbers $(n,l,m)$ of the corresponding state. They are defined by
\begin{equation}\label{eq:shannon_rho}
S \left[ \rho \right]= - \int \rho(\vec{r}) \ln \rho(\vec{r}) d\vec{r},
\end{equation}
and
\begin{equation}\label{eq:shannon_gamma}
S\left[ \gamma \right]= - \int \gamma(\vec{p}) \ln \gamma(\vec{p}) d\vec{p}.
\end{equation}

From Eqs. (\ref{eq:rho}) and (\ref{eq:shannon_rho}), and (\ref{eq:gamma}) and (\ref{eq:shannon_gamma}), one obtains that the Shannon entropy has the value
\begin{equation}\label{eq:shannon_rho_res}
S \left[ \rho \right]= A_{n,l}+\frac{1}{2 n} E_1\left[ \tilde{{\cal{L}}}^{2l+1}_{n-l-1} \right]+S\left(Y_{l,m}\right)-3 \ln Z,
\end{equation}
and
\begin{equation}\label{eq:shannon_gamma_res}
S\left[ \gamma \right] = B_{n,l}+ E_0 \left[ \tilde{{\cal{C}}}^{l+1}_{n-l-1} \right]+S \left(Y_{l,m}\right)+3 \ln Z,
\end{equation}
in position and momentum spaces, respectively. The constants $A_{n,l}$ and $B_{n,l}$ have the expresions
\begin{equation*}
A_{n,l} = -\ln \left( \frac{4}{n^4} \right) +\frac{3 n^2-l (l+1)}{n} - 2l \left[\frac{2n-2l-1}{2 n}+\psi (n+l+1) \right],
\end{equation*}
and
\begin{equation*}
B_{n,l}=-\ln \frac{n^3}{2^{2l+4}}-(2l+4) \left[ \psi(n+l+1)-\psi (n) \right]+\frac{l+2}{n}-4 \left[1-\frac{2 n (2l+1)}{4n^2-1} \right].
\end{equation*}

The symbols $E_{k} \left[ y_{n}\right]$, $k=0$ and $1$, denote the following entropic integrals of the polynomials $y_{n} (x)$ orthonormal with respect to the weight function $\omega(x)$, $a<x<b$:
\begin{equation*}
E_k\left[ y_{n}\right]= - \int_{a}^{b} x^k \omega (x) y_{n}^2 (x) \ln y_{n}^2(x) dx.
\end{equation*}

These integrals cannot be determined analytically except for the two extreme cases with small and large $n$ in the Laguerre case and $n=0$, $n=1$ and $n\gg 1$ in the Gegenbauer case. The calculation for $n\gg 1$ is a formidable task which can be tackled in the Laguerre \cite{dehesa:jmp98} and Gegenbauer \cite{aptekarev:rassm95, aptekarev:jmp94} cases by means of some highbrow methods of approximation theory. We have obtained
\begin{align}
\nonumber
E_1\left[{\tilde{\cal{L}}}^{\alpha}_{n}\right]&:=-\int^{\infty}_{0} x \omega_\alpha (x) \left[ \tilde{\cal{L}}^{\alpha}_{n}(x) \right]^2 \ln\left[ \tilde{\cal{L}}^{\alpha}_{n}(x) \right]^2 dx\\\label{eq:entropic_L}
&= -6 n^2+ 2 (\alpha+1)n \ln n + 2n \left( \ln (2 \pi)-2 \alpha-4 \right)+o(n),
\end{align}
and
\begin{align}\nonumber
E_0\left[{\tilde{\cal{C}}}^{\alpha}_{n}\right]&:=-\int^{+1}_{-1} \omega^{*}_\alpha (x) \left[ \tilde{\cal{C}}^{\alpha}_{n}(x) \right]^2 \ln\left[ \tilde{\cal{C}}^{\alpha}_{n}(x) \right]^2 dx\\\label{eq:entropic_C}
&= \ln \pi+(1-2 \alpha) \ln 2-1+o(1),
\end{align}
for $n \gg 1 $, respectively. The remaining symbol $S(Y_{l,m})$ in Eqs. (\ref{eq:shannon_rho_res})-(\ref{eq:shannon_gamma_res}) denotes the angular Shannon entropy given by \cite{yanez:jmp99}
\begin{align}\label{eq:shannon_Y}
S \left(Y_{l,m}\right)& = - \int \left| Y_{l,m} (\theta, \varphi) \right|^2 \ln  \left| Y_{l,m} (\theta, \varphi) \right|^2 \sin \theta d\theta d \varphi \nonumber \\
& = D_{l,m} + E \left[ \tilde{{\cal{C}}}^{m+\frac{1}{2}}_{l-m} \right],
\end{align}
with
\begin{equation*}
D_{l,m}=\ln 2\pi - 2 m \left[ \psi (l+m+1)- \psi \left(l+\frac{1}{2} \right) - \ln 2 - \frac{1}{2l+1} \right].
\end{equation*}

Then, from Eqs. (\ref{eq:shannon_rho_res}) and (\ref{eq:entropic_L}), and (\ref{eq:shannon_gamma_res}) and (\ref{eq:entropic_C}), one obtains that the Shannon information entropies of the Rydberg states $(n,l,m)$, $n \rightarrow \infty$ and ($l,m$) fixed, have the following expressions
\begin{equation}\label{eq:S_rho_res_n}
S\left[ \rho \right] = 6 \ln n- \ln 2 +\ln \pi-3 \ln Z + S(Y_{l,m}) + o (1),
\end{equation}
\begin{equation}\label{eq:S_gamma_res_n}
S\left[ \gamma \right] = -3 \ln n + 3 \ln 2 +\ln \pi-5 + 3 \ln Z + S(Y_{l,m}) + o (1),
\end{equation}
in position and momentum spaces, respectively. The angular Shannon entropy $S(Y_{l,m})$ is given by Eq. (\ref{eq:shannon_Y}). This integral cannot be analytically calculated for arbitrary $(l,m)$ values, but it is possible to do it for some particular cases. Indeed, for states with $\left| m \right|=l$ one has
\begin{equation}\label{eq:shannon_Y_ll}
S\left( Y_{l,l} \right)= \ln \left( \frac{2^{2l+1} \pi^{\frac{3}{2}} l!}{\Gamma \left(l+\frac{3}{2} \right)} \right) -2 l \left[ \psi \left( 2l+1 \right)- \psi \left( l+\frac{1}{2}\right) - \frac{1}{2l+1}\right],
\end{equation}
so that, in particular for $ns-$states (i.e. when $l=m=0$) one has $S(Y_{0,0})= \ln 4\pi$. Then, from Eqs. (\ref{eq:S_rho_res_n}) and (\ref{eq:S_gamma_res_n}), we can obtain the values:
\begin{equation*}
S\left[ \rho \right] = 6 \ln n+ \ln 2 + 2\ln \pi-3 \ln Z  + o (1),
\end{equation*}
\begin{equation*}
S\left[ \gamma \right] = -3 \ln n + 5 \ln 2 +2 \ln \pi-5 + 3 \ln Z  + o (1),
\end{equation*}
for the position and momentum Shannon entropy of Rydberg $ns-$states (also called linear states). On the other hand, when $\left| m \right|=l-1$ then one has
\begin{eqnarray}
S \left(Y_{l,l-1}\right)&=& \ln \left( \frac{2^{2l} \pi^{\frac{3}{2}} (l-1)!}{\Gamma \left(l+\frac{3}{2} \right)} \right) -2+ \gamma_{\text{EM}} +\psi \left(l+\frac{3}{2} \right)\nonumber\\
&& -2(l-1)\left[ \psi \left( 2l \right)- \psi \left( l+\frac{1}{2}\right) - \frac{1}{2l+1}\right],
\label{eq:shannon_Y_ll-1}
\end{eqnarray}
where $\gamma_{\text{EM}}=0.57721$ is the Euler-Mascheroni constant.
Taking into account the asymptotic behavior of the gamma
\begin{equation}\label{eq:asymp_gamma}
\Gamma(x) = \sqrt{2 \pi} e^{-x} x^{x-1/2} \left(1+\frac{1}{12x} + O \left( x^{-2} \right) \right); \qquad \qquad \text{for} \quad x \rightarrow \infty,
\end{equation}
and digamma
\begin{equation}\label{eq:asymp_digamma}
\psi(x) = \ln x -\frac{1}{2x}-\frac{1}{12x^2} + O \left( x^{-4} \right); \qquad \qquad \text{for} \quad x \rightarrow \infty,
\end{equation}
functions,
one obtains from Eqs. (\ref{eq:shannon_Y_ll}) and (\ref{eq:shannon_Y_ll-1}) the asymptotics of the angular Shannon entropies
\begin{equation*}
S \left(Y_{l,l}\right)= \ln2 +\frac{3}{2} \ln \pi + \frac{1}{2} - \frac{1}{2} \ln l + O \left( l^{-1} \right),
\end{equation*}
and 
\begin{equation*}
S \left(Y_{l,l-1}\right) = 2 \ln2 +\frac{3}{2}  \ln \pi + \gamma_{\text{EM}} - \frac{1}{2} - \frac{1}{2} \ln l + O \left( l^{-1} \right).
\end{equation*}

In addition, from Eqs. (\ref{eq:entropic_C}) and (\ref{eq:shannon_Y}), when  $m=0$, one gets the asymptotics
\begin{equation*} 
S\left(Y_{l,0}\right)= \ln 2 + 2 \ln \pi -1 +o(1) \qquad \text{for} \quad l \gg 0.
\end{equation*}

Finally, let us remark that the entropic sum obtained from Eqs. (\ref{eq:S_rho_res_n}) and (\ref{eq:S_gamma_res_n}) fulfils not only the general entropic uncertainty relation given by Eq. (\ref{eq:S_rhoS_gamma}) but also the improved corresponding relation of central potentials \cite{rudnicki_12} recently derived, what is a further checking of our results.

\section{Position and momentum Fisher informations}
\label{sec:fisher}

The Fisher information of the probability density $\rho \left(\vec{r}\right)$ is defined by
\begin{equation}
\label{eq:fisher}
I \left[ \rho \right]= \int \frac{\left| \vec{\nabla} \rho \left(\vec{r}\right)\right|^2}{\rho \left(\vec{r}\right)} d\vec{r}.
\end{equation}

This quantity, which is the translationally invariant version of the parameter-dependent Fisher information originally introduced by R. E. Fisher in the statistical estimation theory \cite{fisher:pcps25}, has been applied in a great number of scientific and technological fields. Contrary to the remaining spreading measures, the Fisher information is a local quantity because it is very sensitive to the fluctuations or irregularities of the density. Note that it is a functional of the gradient of $\rho \left( \vec{r}\right)$, so that it nicely grasps the highly oscillatorial behavior of the Rydberg wavefunctions. The higher it is, the more localized is the density, the smaller is the uncertainty and the higher is the accuracy in estimating the localization of the electron. Moreover, the Fisher information is closely related to the kinetic and Weizs\"acker energies.

The explicit expressions for the position and momentum Fisher information of a hydrogenic state ($n,l,m$) can be obtained either algebraically by means of the numerous properties of the orthogonal polynomials which control its wavefunctions or by use of the relations
\begin{align}
I \left[\rho \right] &= 4 \left\langle p^2 \right\rangle - 2 \left| m \right| (2l+1) \left\langle r^{-2} \right\rangle, \\ 
I \left[\gamma \right] &= 4 \left\langle r^2 \right\rangle - 2 \left| m \right| (2l+1) \left\langle p^{-2} \right\rangle,
\end{align}
together with the known expressions of the radial expectations values ($\left\langle r^2 \right\rangle$, $\left\langle r^{-2} \right\rangle$, $\left\langle p^2 \right\rangle$, $\left\langle p^{-2} \right\rangle$) in the two reciprocal spaces. One obtains the values
\begin{equation}\label{eq:I_rho_def}
I \left[ \rho \right]= \frac{4 Z^2}{n^3} \left(n- \left| m \right| \right),
\end{equation}
and 
\begin{equation}\label{eq:I_gamma_def}
I \left[ \gamma \right]= \frac{2 n^2}{Z^2} \left[ 5n^2- 3l (l+1)- \left| m \right| (8n-6l-3)+1 \right],
\end{equation}
for the Fisher information of the hydrogenic state $(n,l,m)$ in position and momentum spaces, respectively.

Therefore, the Fisher informations of the Rydberg states $(n,l,m)$, $n\rightarrow \infty$ and fixed $(l,m)$ have the values
\begin{equation*}
I\left[ \rho \right] = \left(\frac{2 Z}{n} \right)^2 + O \left( \frac{1}{n^3} \right),
\end{equation*}
\begin{equation*}
I\left[ \gamma \right] = \frac{10}{Z^2} n^4 + O (n^3),
\end{equation*}
in the two reciprocal spaces. It is worth remarking that the net uncertainty product
\begin{equation*}
I \left[\rho \right] \times I \left[\gamma \right] \geq 40 n^2 +O(n),
\end{equation*}
for these Rydberg states, certainly fulfils the uncertainty relation (\ref{eq:I_rhoI_gamma}). Moreover, it is interesting to note that the position and momentum Fisher information scale as the energy and the geometric cross section of the Rydberg state, that is as $n^{-2}$ and $n^4$, respectively.

\section{Complexity measures}

In this Section we find the position and momentum complexity measures of Cr\' amer-Rao, Fisher-Shannon and LMC types of the probability density of the Rydberg states $(n,l,m)$ with $n\rightarrow \infty$ and $(l,m)$ fixed.
The Cr\' amer-Rao complexity measure \cite{dehesa:jcam06,dembo:ieetit91,ANT} in position space is given by
\begin{equation}
\label{eq:cramerho}
C_{CR}[\rho]=V[\rho]\times I[\rho],
\end{equation}
where $V[\rho]=\left\langle r^{2} \right\rangle -|\langle \vec{r}\rangle|^2$ is the well-known variance, and $I[\rho]$ denotes the Fisher information (\ref{eq:fisher}) discussed previously.
Moreover, since $|\langle \vec{r}\rangle|^2 =0$ for any state, we have that 
\[
V[\rho] =\left\langle r^{2} \right\rangle = \frac{n^{2}}{2Z^{2}}\left[5n^{2}-3l(l+1)+1\right]
\] 
Then, the exact value of the Cr\' amer-Rao complexity for any hydrogenic state $(n,l,m)$ of the atom in position space is 
\begin{equation}
C_{CR}[\rho]=\frac{2}{n}(5n^2-3l(l+1)+1)(n-|m|)
\end{equation}
Similarly, taking into account that $\left\langle p^{2} \right\rangle =\frac{Z^{2}}{n^{2}}$ and the value (\ref{eq:I_gamma_def}) for the momentum Fisher information $I[\gamma]$ we obtain in a straightforward manner that 
\begin{equation}
C_{CR}[\gamma]=2 (5 n^2 - 3 l (l+1) - |m| (8 n - 6 l - 3) + 1),
\end{equation}
is the exact value for the momentum Cr\' amer-Rao complexity of any hydrogenic state $(n,l,m)$. From the two previous expressions, it is interesting to realize that for $l$ fixed and large $n$, one has 
\begin{equation}
\label{eq:cr-asymp}
C_{CR}[\rho]=10n^2+O(n) \quad \text{and} \quad C_{CR}[\gamma]=10n^2+O(n),
\end{equation}
which indicates that the Cr\' amer-Rao complexity of Rydberg states with $l$ fixed have the same leading behaviour.\\
Let us now examine the Fisher-Shannon complexity \cite{ANG,ROM}, which is defined by
\begin{equation}
\label{eq:crho-asymp}
C_{FS}[\rho]=I[\rho]\times \frac{1}{2\pi e}e^{\frac23 S[\rho]},
\end{equation}
in position space.
This quantity measures the combined balance of the gradient content of the density $\rho(\vec{r})$ as given by the Fisher information $I[\rho]$, and the spatial spreading as given by an exponential of the Shannon entropy $S[\rho]$.
Thus, the Fisher-Shannon complexity grasps a two-fold facet of the density of localization-delocalization or oscillatory-spreading character. The two components of this complexity, $S[\rho]$ and $I[\rho]$, have been discussed in Sections III and IV, respectively. While the Fisher information is exactly known for any general hydrogenic state, we only know the leading term of the Shannon entropy for the Rydberg states. Then, from (\ref{eq:S_rho_res_n}) and (\ref{eq:I_gamma_def}) and (\ref{eq:crho-asymp}) one has the value
\[
C_{FS}[\rho]=\left(\frac{2}{\pi}\right)^\frac13 e^{-1+\frac23 S(Y_{l,m})} n^2 + o(n^2),
\]
for the position Fisher-Shannon complexity of a Rydberg state $(n,l,m)$ with $l$ fixed. Here, $S(Y_{l,m})$ denotes the angular Shannon entropy which can be calculated either numerically or analytically from (\ref{eq:shannon_Y}) for each pair $(l,m)$; see section III for its analytical expression in some particular cases. \\
The Fisher-Shannon complexity in momentum space, $C_{FS}[\gamma]$, of Rydberg states can be calculated in a similar manner by making profit of the exact value (\ref{eq:I_gamma_def}) of the momentum Fisher information $I[\gamma]$ and the asymptotic value (i.e., large $n$) of the momentum Shannon entropy given by (\ref{eq:S_gamma_res_n}). We found that
\[
C_{FS}[\gamma]=20\pi^{-\frac13}e^{-\frac{13}{3}+\frac23 S(Y_{l,m})} n^2+o(n^2),
\]
when $n\rightarrow \infty$ and $(l,m)$ is fixed, where $S(Y_{l,m})$ denotes the angular part of the Shannon entropy which can be obtained from (\ref{eq:shannon_Y}) for each pair $(l,m)$. It is worth pointing out that the expressions of the leading term of both position and momentum Fisher-Shannon complexities of the Rydberg states can be explicitly given in terms of the quantum numbers for some particular cases; not only for linear states $(l=m=0)$, where $S(Y_{00}=\ln 4\pi)$, but also for the states with $(l,m)=(l,l)$ and $(l,l-1)$ for which the associated angular Shannon entropies $S(Y_{lm})$ have been already given in (\ref{eq:shannon_Y_ll}) and (\ref{eq:shannon_Y_ll-1}), respectively. Notice that we have the values
\begin{eqnarray}
\label{eq:cfs_n00}
C_{FS}[\rho]&\simeq & \sqrt[3]{2^{5}\pi}e^{-1} n^{2} +o(n^2)\\
C_{FS}[\gamma]&\simeq & 5\sqrt[3]{2^{4}\pi}e^{-\frac{13}{3}} n^{2} +o(n^2)
\end{eqnarray}
for the position and momentum Fisher-Shannon complexity measures of the linear Rydberg states $(n,0,0)$.
\\
Finally, let us consider the LMC complexity measure $C_{LMC}[\rho]$ which is given \cite{catalan:pre02} by 
\begin{equation}
\label{eq:lmcrho}
C_{LMC}[\rho]=D[\rho]\times e^{S[\rho]},
\end{equation}
in position space. The symbols $N[\rho] \equiv e^{S[\rho]}$ and $D[\rho] \equiv W_2[\rho]$ denote the Shannon entropy power and the disequilibrium or second-order entropic moment of the density
\begin{equation}
\label{eq:disequi_rho}
D[\rho] = \left\langle \rho \right\rangle = \int_{\mathbb{R}^{3}} (\rho(\vec{r}))^{2}\, d\vec{r},
\end{equation}
respectively. Then, this complexity measure quantifies the combined balance of the average height (as given by the disequilibrium) and the total extent of the density (as given by the Shannon entropy power). To calculate this quantity for Rydberg states we use the asymptotic value (i.e. large $n$) of the Shannon entropy $S[\rho]$ given by (\ref{eq:S_rho_res_n}), and the exact value of the disequilibrium $D[\rho]$ which is given \cite{XX} by
\begin{eqnarray}
\label{eq:disequilibrium}
D[\rho] &=& \frac{(2l+1)^{2}}{2^{4n}\pi n^{5}}\sum_{k=0}^{n_{r}}\left( \begin{array}{c}
  														2n_{r} - 2k   \\
  														n_{r} - k
  														  \end{array} \right)^{2} \nonumber\\
	     &\times&\frac{(k+1)_{k}}{k!}\frac{\Gamma(4l+2k+3)}{\Gamma^{2}(2l+k+2)} \nonumber \\
	     &\times&\sum_{l'=0}^{2l}(2l' +1)\left( \begin{array}{ccc}
  								   l & l & l'   \\
  								   0 & 0 & 0 
  								   \end{array} \right)^{2} \left( \begin{array}{ccc}
  								   l & l & l'   \\
  								   m & m & -2m 
  								   \end{array} \right)^{2}
\end{eqnarray}
with $n_{r} =n-l-1$. Then the use of (\ref{eq:S_rho_res_n}) and (\ref{eq:disequilibrium}) allows us to obtain the position LMC complexity measure for Rydberg states. The resulting expression can be simplified by finding the value of $D[\rho]$ in the limiting case $n\rightarrow \infty$ from (\ref{eq:disequilibrium}), whose achievement is left to the reader. Working in a similar manner one can obtain the LMC complexity measure in momentum space for all Rydberg states by means of the expression (\ref{eq:S_gamma_res_n}) for the momentum Shannon entropy $S[\gamma]$ and the value of the momentum disequilibrium $D[\gamma] = \langle \gamma\rangle$. \\
Thus, the evaluation of the position and momentum LMC complexity measures remains to be an open problem not only for general hydrogenic states (because the values of the position and momentum Shannon entropies and the momentum disequilibrium are unknown) but also for Rydberg states (because the limiting case $n\rightarrow \infty$ of the position and momentum disequilibrium are also unknown). Nevertheless, we show in the next section that the leading term of the LMC complexity measures can be obtained for the special case of circular and quasicircular states in both position and momentum states.

\section{Application to circular and quasicircular states}
\label{sec:circular}

In this Section we calculate the position and momentum spreading measures of the circular $(l=\left| m \right|=n-1)$ and quasicircular $(l=\left| m \right|=n-2)$ states of Rydberg atoms by use of the theoretical methodology obtained in the previous sections. These measures quantify in various complementary ways the high anisotropy of the electron density distribution of the circular and quasicircular Rydberg atoms. These systems are ideal semiclassical objects at the frontier of classical and quantum physics since the electron distribution in position space has the shape of a thin torus along the Bohr circular orbit of radius $\sim n^2$ atomic units. Moreover, they are also important because they are experimentally accessible and relatively long lived \cite{chen:jpb93, delande:el88, liang:pra86, hulet:prl83}. As the electron distribution does not penetrate the atomic core, any Rydberg atom in a circular or a quasicircular state is a quasi-hydrogenic system.

\subsection{Circular states: $\mathbf{(n, n-1,n-1)}$}

When the orbital angular momentum, which describes the state of the system, has the maximum value $l = n-1$, the electron orbit approaches a circular path with radius $r = \left\langle r \right\rangle$. In this case, for $l=m=n-1$, the probability densities are given by
\begin{equation}
\label{eq:circrho}
\rho(\vec{r})= \frac{4 Z^3}{n^4 \Gamma(2n)} {\tilde{r}}^{2n-2} e^{- \tilde{r}}  \left| Y_{n-1,n-1}(\theta,\varphi) \right|^2,
\end{equation}
and
\begin{equation}
\label{eq:circgam}
\gamma(\vec{p})= \frac{2^{4n+1} n^4 }{\pi Z^3 } \frac{\Gamma^2(n)}{\Gamma(2n)} \frac{(n \tilde{p})^{2n-2}}{(1+n^2 {\tilde{p}}^2)^{2n+2}}  \left| Y_{n-1,n-1}(\theta,\varphi) \right|^2,
\end{equation}
in position and momentum spaces respectively.

For these states, the position and momentum power moments can be determined in an exact form in position space:
\[
\langle r^\alpha\rangle = \left(\frac{n}{2Z}\right)^\alpha \frac{\Gamma(2n+\alpha+1)}{\Gamma(2n+1)},\quad \alpha > -2n-1,
\]
and in momentum space:
\[
\langle p^\alpha\rangle = \left(\frac{Z}{n}\right)^\alpha
\frac{\Gamma\left(\frac{3-a}{2}+n\right)\Gamma\left(\frac{1+a}{2}+n\right)}{\Gamma\left(\frac{1}{2}+n\right)\Gamma\left(\frac{3}{2}+n\right)}, \quad  -2n-1 < \alpha < 2n+3.
\]
Taking into account the asymptotic behaviour of the gamma function (\ref{eq:asymp_gamma}),
the following asymptotics of these expectation values are obtained for large values of $n$:
\begin{equation}
\label{eq:alpharad}
\langle r^\alpha\rangle = \left(\frac{n^2}{Z}\right)^\alpha \left( 1+ O\left(\frac{1}{n}\right)\right),
\end{equation}
\begin{equation}
\label{eq:alphamom}
\langle p^\alpha\rangle = \left(\frac{Z}{n}\right)^\alpha \left( 1+ O\left(\frac{1}{n}\right)\right).
\end{equation}
The logarithmic expectations values turn out to be given by
\[
\langle \ln r\rangle =\ln\left(\frac{n}{2Z}\right)+\psi(2n+1),
\qquad
\langle \ln p\rangle =-\ln\left(\frac{n}{Z}\right)-\frac{1}{2n+1},
\]
and their asymptotic behaviours are
\[
\langle \ln r\rangle = 2\ln n -\ln Z+ O\left(\frac{1}{n}\right),
\qquad
\langle\ln p\rangle = -\ln n+\ln Z+ O\left(\frac{1}{n}\right),
\]
where we have taken into account the asymptotics of the digamma function $\psi(x)$ (\ref{eq:asymp_digamma}).

The Shannon entropies, given by Eq. (\ref{eq:shannon_rho_res}) and (\ref{eq:shannon_gamma_res}) can be analytically determined for the circular states:
\begin{align}\label{eq:shannon_rho_cir}
 S\left [\rho_{n,n-1,n-1} \right] & = A_{n,n-1}+ \frac{1}{2n} E_1\left[{\tilde{\cal{L}}}^{2n-1}_{0}\right]+ S \left(Y_{n-1,n-1} \right) -3 \ln Z \nonumber \\
&= \ln \left( \frac{\pi n^4 }{Z^3} \right)+2 \ln \left[\Gamma(n) \right]+2n+1-2(n-1) \left[\frac{1}{2 n}+ \psi(n) \right],
\end{align}
in position space, and
\begin{align}\label{eq:shannon_gamma_cir}
  S\left [\gamma_{n,n-1,n-1} \right] & = B_{n,n-1}+  E_0 \left[{\tilde{\cal{C}}}^{n}_{0}\right]+ S \left(Y_{n-1,n-1} \right) +3 \ln Z \nonumber \\
&= \ln \left( \frac{2^5 \pi^2 Z^3}{ n^4 }\right)+ \frac{n+1}{n}- \frac{4}{2n-1} +4 \left[ \psi(n)-\psi(2n) \right],
\end{align}
in momentum space. In order to obtain the Shannon entropies for the Rydberg circular states, one has to calculate the limit $n \rightarrow \infty$ in Eqs. (\ref{eq:shannon_rho_cir}) and (\ref{eq:shannon_gamma_cir}). 

Taking into account the asymptotic behaviour of the gamma and digamma functions, and the Taylor series of the logaritmic function
\begin{equation*}
\ln (1+x) = x-\frac{x^2}{2}+\frac{x^3}{3}+ O \left( x^{4} \right); \qquad \qquad \text{for} \quad x\rightarrow 0,
\end{equation*}
one can obtain
\begin{equation*}
S \left[ \rho_{n,n-1,n-1} \right] = 5 \ln n+ 1 + \ln \left(\frac{2 \pi^2}{Z^3} \right)+\frac{1}{3n}+ O\left( \frac{1}{n^2} \right),
\end{equation*}
and
\begin{equation*}
S \left[ \gamma_{n,n-1,n-1} \right] = -4 \ln n- \frac{2}{n} +1 + \ln \left(2 \pi^2 Z^3 \right)+ O\left( \frac{1}{n^2}\right),
\end{equation*}
for the Shannon entropy of the Rydberg circular states $(n \rightarrow \infty)$ in position and momentum spaces, respectively.

The Fisher information can also be exactly determined for circular states. Taking Eqs. (\ref{eq:I_rho_def}) and (\ref{eq:I_gamma_def}) with $|m|=l=n-1$, we have
\begin{equation}
\label{eq:asympfishc}
I[\rho_{n,n-1,n-1}]=\frac{4Z^2}{n^3},\quad \text{and} \quad
I[\gamma_{n,n-1,n-1}]=\frac{4}{Z^2}n^2(n+2),
\end{equation}
for the position and momentum spaces, respectively. \\
According to the previous results, all the complexity measures presented in the previous section can be exactly evaluated both in position and momentum space. Let us show their corresponding values in an explicit manner. 

According to (\ref{eq:circrho}) and (\ref{eq:circgam}), or directly from (\ref{eq:cramerho}),(\ref{eq:alpharad}) and (\ref{eq:alphamom}) with $\alpha =2$, and (\ref{eq:asympfishc}), one obtains that Cr\' amer-Rao complexities of the circular hydrogenic states have the values
\[
C_{CR}[\rho_{n,n-1,n-1}]=4n+6+\frac{2}{n},
\]
and
\[
C_{CR}[\gamma_{n,n-1,n-1}]=4n+8,
\]
in the position and momentum spaces, respectively. Notice that for Rydberg states, these quantities have the same leading term.

According to (\ref{eq:crho-asymp}), (\ref{eq:shannon_rho_cir}), (\ref{eq:shannon_gamma_cir}) and (\ref{eq:asympfishc}) we have obtained that the Fisher-Shannon complexities of the circular hydrogenic states have the expressions
\[
C_{FS}[\rho_{n,n-1,n-1}]=2(\pi n)^{-\frac13} \left(\Gamma(n)\right)^\frac43 \exp\left[
\frac23 \left(2n-\frac12 -2(n-1)\left(\frac{1}{2n}+\psi(n)\right)\right)
\right]
\]
and
\[
C_{FS}[\gamma_{n,n-1,n-1}]=16(2\pi n)^\frac13 (n+2)\exp\left[
\frac23 \left(
-\frac12 +\frac{1}{n}-\frac{4}{2n-1}+4\left(\psi(n)-\psi(2n)\right)
\right)
\right]
\]
in position and momentum spaces, respectively. Taking into account the asymptotic behaviour of the gamma and digamma functions given by (\ref{eq:asymp_gamma}) and (\ref{eq:asymp_digamma}), we have found that the leading term of Fisher-Shannon complexity of circular Rydberg states has the same value:
\[
C_{FS}[\rho_{n,n-1,n-1}]=C_{FS}[\gamma_{n,n-1,n-1}] =2^\frac53 \left(\frac{\pi}{e}\right)^\frac13 n^\frac13 + O\left(n^{-\frac23}\right)
\]

in the position and momentum spaces, respectively.\\
Finally, according to its definition (\ref{eq:lmcrho}), to evaluate the LMC complexity measures of the circular hydrogenic states we have to calculate both the Shannon entropy (see (\ref{eq:shannon_rho}) and (\ref{eq:shannon_gamma})) and the disequilibrium (see (\ref{eq:disequilibrium})) in the two reciprocal spaces directly from the corresponding probability densities given by (\ref{eq:circrho}) and (\ref{eq:circgam}). The Shannon entropies are given by (\ref{eq:shannon_rho_cir}) and (\ref{eq:shannon_gamma_cir}), and the disequilibrium has the expressions  
\[
D[\rho_{n,n-1,n-1}]=\frac{(2n-1)\left(\Gamma\left(n-\frac12\right)\right)^2Z^3}{8\pi^2 n^5\left(\Gamma(n)\right)^2},
\]
and
\[
D[\gamma_{n,n-1,n-1}]=\frac{n^4(4n+5)(4n+7)Z^{-3}}{16\pi^2(4n^2-1)},
\]
in the position and momentum spaces, respectively. With these values and those of the Shannon-entropy expressions (\ref{eq:shannon_rho_cir}) and (\ref{eq:shannon_gamma_cir}), we obtain that the LMC complexity measures of circular hydrogenic states have the values
\[
C_{LMC}[\rho_{n,n-1,n-1}]=
\frac{(2n-1)\left(\Gamma\left(n-\frac12\right)\right)^2}{8\pi n}
\exp\left[
2n+1-2(n-1)\left(\frac{1}{2n}+\psi(n)\right)
\right],
\]
and
\[
C_{LMC}[\gamma_{n,n-1,n-1}]=
\frac{2(4n+5)(4n+7)}{4n^2-1}
\exp\left[
1+\frac{1}{n}-\frac{4}{2n-1}+4\left(
\psi(n)-\psi(2n)
\right)
\right]
\]
for the position and momentum spaces, respectively. The asymptotics ($n\rightarrow \infty$) of these quantities yield the expressions
\[
C_{LMC}[\rho_{n,n-1,n-1}]=\frac{e}{2}+\frac{7e}{24n}+O\left(n^{-2}\right),
\]
and
\[
C_{LMC}[\gamma_{n,n-1,n-1}]=\frac{e}{2}+\frac{e}{2n}+O\left(n^{-2}\right),
\]
for the position and momentum LMC complexities of Rydberg circular states in the position and momentum spaces, respectively. As in the previous complexity measures, the main term of the asymptotics is identical for both densities.

\subsection{Quasicircular states: $\mathbf{(n, n-2,n-2)}$}

Let us consider the Rydberg states with orbital quantum numbers $l=m=n-2$.
In this case, the probability densities are given by
\begin{equation*}
\rho(\vec{r})= \frac{4 Z^3}{n^4 \Gamma(2n-1)} {\tilde{r}}^{2n-4} e^{- \tilde{r}} 
(2n-2+\tilde{r})^2
\left| Y_{n-2,n-2}(\theta,\varphi) \right|^2,
\end{equation*}
and
\begin{equation*}
\gamma(\vec{p})= \frac{2^{4n-1} n^4(n-1)^2 }{\pi Z^3 } \frac{\Gamma^2(n-1)}{\Gamma(2n-1)} \frac{(n \tilde{p})^{2n-4}(1-n^2 \tilde{p}^2)^2}{(1+n^2 {\tilde{p}}^2)^{2n+2}}  \left| Y_{n-2,n-2}(\theta,\varphi) \right|^2,
\end{equation*}
in position and momentum spaces respectively.

For these states, the position and momentum power moments can be determined in an exact form in position space:
\[
\langle r^\alpha\rangle =\left(\frac{n}{2Z}\right)^\alpha \frac{2n+3\alpha +\alpha^2}{2n} \frac{\Gamma(2n+\alpha-1)}{\Gamma(2n-1)},\quad \alpha >-2n+1,
\]
and in momentum space:
\[
\langle p^\alpha\rangle = \left(\frac{Z}{n}\right)^\alpha
\frac{2n+(\alpha-1)^2}{2}
\frac{\Gamma\left(n+\frac{a-1}{2}\right)\Gamma\left(n+\frac{1-a}{2}\right)}{\Gamma\left(n-\frac{1}{2}\right)\Gamma\left(n+\frac{3}{2}\right)},\quad -2n+1<\alpha <2n+1.
\]
Taking into account the asymptotic behaviour of the gamma function (\ref{eq:asymp_gamma})
the asymptotics of these expectation values are obtained for large values of $n$:
\[
\langle r^\alpha\rangle = \left(\frac{n^2}{Z}\right)^\alpha \left( 1+ O\left(\frac{1}{n}\right)\right),
\]
\[
\langle p^\alpha\rangle = \left(\frac{Z}{n}\right)^\alpha \left( 1+ O\left(\frac{1}{n}\right)\right).
\]
The logarithmic expectations values turn out to be given by
\[
\langle \ln r\rangle =\ln\left(\frac{n}{2Z}\right)+\frac{3}{2n}+\psi(2n-1),
\qquad
\langle \ln p\rangle =-\ln\left(\frac{n}{Z}\right)-\frac{6n-1}{4n^2-1},
\]
and their asymptotic behaviours are
\[
\langle \ln r\rangle = 2\ln n -\ln Z+ O\left(\frac{1}{n}\right),
\qquad
\langle\ln p\rangle = -\ln n+\ln Z+ O\left(\frac{1}{n}\right),
\]
where we have taken into account the asymptotics of the digamma function (\ref{eq:asymp_digamma}).

In position space, the Shannon entropy (\ref{eq:shannon_rho_res}) for the quasicircular states has the form
\begin{align}\label{eq:shannon_rho_quasicir}
S\left [\rho_{n,n-2,n-2} \right] & = A_{n,n-2}+ \frac{1}{2n} E_1\left[{\tilde{\cal{L}}}^{2n-3}_{1}\right]+ S \left(Y_{n-2,n-2} \right) -3 \ln Z \nonumber \\
&= \ln \left( \frac{2 \pi }{Z^3} \right)+3 \ln(n)- \ln \left(1- \frac{1}{n} \right)+2n +1+\frac{4}{n}-\frac{1}{n-1} \nonumber \\
&\quad +2 \ln \left[\Gamma(n) \right]-2(n-2) \psi(n)-\frac{1}{2n \Gamma(2n-1)} I_n,
\end{align}
where the integral $I_n$ is defined as
\begin{equation}
\label{eq:rydbergint}
I_n= \int_{0}^{\infty} x^{2n-2} e^{-x} (2n-2-x)^2 \ln (2n-2-x)^2 dx.
\end{equation}
This integral can be analytically expressed (see Appendix) in terms of the generalized hypergeometric series ${}_2 F_{2}(a_{1},a_{2};b_{1},b_{2};z)$ and ${}_1 F_{1}(a;b;z)$, from which one can heuristically obtain that
\begin{equation}
-\frac{1}{2n\Gamma(2n-1)}I_{n} = -\ln n + \gamma_{_{EM}} -2 + o(1)
\end{equation}

Then, taking into account all the terms in Eq. (\ref{eq:shannon_rho_quasicir}), the asymptotics of the Shannon entropy of the quasicircular states in the position space is given by
\begin{equation}
\label{eq:asympshanrad}
S\left [\rho_{n,n-2,n-2} \right] = 5\ln n +\ln\frac{(2\pi)^{2}}{Z^{3}} + \gamma_{_{EM}} + o(1) .
\end{equation}

For the momentum space, the expression (\ref{eq:shannon_gamma_res}) for quasicircular states yields
\begin{align}\label{eq:shannon_gamma_quasicir}
S \left [\gamma_{n,n-2,n-2} \right] & = B_{n,n-2}+ E_0\left[{\tilde{\cal{C}}}^{n-1}_{1}\right]+ S \left(Y_{n-2,n-2} \right) +3 \ln Z \nonumber \\
&= \ln \left( 2^6 \pi^2 Z^3 \right)-3 \ln (n)+ \ln \left(1- \frac{1}{n} \right) +2 - \frac{1}{n-1} - \frac{8}{2n+1}\nonumber\\
&\quad +4 \left[ \psi(n)-\psi(2n) \right]- \frac{n}{2^{3-2n} \pi} \frac{\Gamma^2(n-1)}{\Gamma(2n-1)}{\tilde{I}_n},
\end{align}
where the integral $\tilde{I_n}$ is given by
\begin{eqnarray}
\label{eq:tildeIint}
\tilde{I}_n &=& \int_{-1}^{1} (1-x^2)^{n-3/2} 4 (n-1)^2 x^2 \ln \left( 4 (n-1)^2 x^2\right) dx. \nonumber\\
&=& 2\sqrt{\pi}(n-1)^2\frac{\Gamma\left(n-\frac12\right)}{\Gamma\left(n+1\right)}\left[\ln(4(n-1)^2)-\psi(n)-\gamma_{\text{EM}}+2(1-\ln 2)-\frac{1}{n}\right],
\end{eqnarray}
where Eq. (2.6.7.2) of Volume 1 of \cite{prudnikov_86} and Eqs. (5.4.2) and (5.4.14) of \cite{olver_10} have been used. With this result, the asymptotics of the last term of Eq. (\ref{eq:shannon_gamma_quasicir}) is
\[
- \frac{n}{2^{3-2n} \pi} \frac{\Gamma^2(n-1)}{\Gamma(2n-1)}{\tilde{I}_n}
=
-\ln n+\gamma_{\text{EM}}-2+\frac{5}{2n}+O\left(n^{-\frac32}\right).
\]
Thus, the asymptotics of the Shannon entropy of quasicircular states for large values of $n$ is
\begin{equation}
\label{eq:asympshan}
S \left[\gamma_{n,n-2,n-2} \right]
=-4\ln n +\ln(4\pi^2 Z^3)+\gamma_{\text{EM}}-\frac{9}{2n}+O\left(n^{-\frac32}\right).
\end{equation}

The Fisher information can also be exactly determined for quasicircular states. Taking Eqs. (\ref{eq:I_rho_def}) and (\ref{eq:I_gamma_def}) with $|m|=l=n-2$, we have
\begin{equation}
\label{eq:asympfish}
I[\rho_{n,n-2,n-2}]=\frac{8Z^2}{n^3},\quad \text{and} \quad
I[\gamma_{n,n-2,n-2}]=\frac{2}{Z^2}n^2(4n+13),
\end{equation}
for the position and momentum spaces, respectively. \\
According to the previous results, all the complexity measures presented in the previous section can be exactly evaluated in momentum space, as well as the Cram\'er-Rao complexity in the position space.

The Cram\'er-Rao complexities read
\[
C_{CR}[\rho_{n,n-2,n-2}]=8n+36-\frac{20}{n},
\]
and
\[
C_{CR}[\gamma_{n,n-2,n-2}]=8n+26,
\]
in the position and momentum spaces, respectively. Notice that, analogously to the circular states, the asymptotic behaviours of these quantities have exactly the same main term $8n$.

According to the asymptotic value (\ref{eq:asympshanrad}) of the Shannon entropy in the position space, the asymptotics of the Fisher-Shannon complexity in this space is given by
\[
C_{FS}[\rho_{n,n-2,n-2}]=8(2\pi)^{1/3}e^{-1+\frac{2}{3}\gamma_{_{EM}}}n^{1/3} + o(n^{1/3}).
\]
On the other hand, the Fisher-Shannon complexity in the momentum space can be given in an exact manner by using expressions (\ref{eq:shannon_gamma_quasicir}), (\ref{eq:tildeIint}) and (\ref{eq:asympfish}). The asymptotics of this quantity can be obtained by means of (\ref{eq:asympshan}) and (\ref{eq:asympfish}):
\[
C_{FS}[\gamma_{n,n-2,n-2}]=8 e^{-1+\frac{2}{3}\gamma_{_{EM}}} (2\pi)^\frac13 n^\frac13 +O\left(n^{-\frac23}\right).
\]

The LMC complexity depends on the disequilibrium of the densities associated to the states in both spaces. Taking into account its definition (\ref{eq:disequi_rho}), the values of this quantity are
\[
W_2[\rho_{n,n-2,n-2}]=\frac{2^{2n-9}(3-2n)^2(6n-7)\left(\Gamma\left(n-\frac32\right)\right)^3 Z^3}{\pi^\frac52 n^5 \Gamma(n-1)\Gamma(2n-1)},
\]
and
\begin{equation}
W_2[\gamma_{n,n-2,n-2}]=\frac{3(n-1)n^4(16n(n+13)+275)}{32\pi^2 (2n-3)(2n-1)(2n+1)Z^3},
\label{eq:w2_gamma_quasi}
\end{equation}
in the position and momentum spaces, respectively. However, like the Fisher-Shannon complexity, the LMC complexity also depends on the Shannon entropy, and its asymptotics in the position space can only be given as
\[
C_{LMC}[\rho_{n,n-2,n-2}]=\frac{3}{4}e^{\gamma_{_{EM}}} + o(1)
\]
In momentum space we can obtain an exact expression of the LMC complexity, taking into account expressions (\ref{eq:shannon_gamma_quasicir}), (\ref{eq:tildeIint}) and (\ref{eq:w2_gamma_quasi}). The asymptotics of this quantity can be obtained by means of (\ref{eq:asympshan}), obtaining:
\[
C_{LMC}[\gamma_{n,n-2,n-2}]=
\frac{3e^{\gamma_{_{EM}}}}{4}+\frac{27e^{\gamma_{_{EM}}}}{4n}+O(n^{-2}).
\]

\section{Conclusions}

In this work we first point out the position and momentum radial ($\langle r^\alpha\rangle$, $\langle p^\alpha\rangle$) and logarithmic ($\langle \ln r\rangle$, $\langle \ln p\rangle$) expectations values with arbitrary order of Rydberg atoms beyond the semiclassical and other approximate methods. We clarify that the exact leading term of the position quantities $\langle r^\alpha\rangle$ for all possible values of $\alpha>-2l-3$ are given by Eq. (\ref{eq:r_alpha_res}), except that of $\alpha =-\frac{3}{2}$ (whose determination remains an open problem). A similar task has been done for the momentum expectation values $\langle p^\alpha\rangle$ of Rydberg atoms, where its leading term is only known when $-1<\alpha<3$ out of all possible values $-2l-3<\alpha<2l+5$. In passing, we observe that our results fulfil the known Heisenberg-like ($\langle r^\alpha\rangle \langle p^\alpha\rangle\ge$ constant) and the logarithmic ($\langle \ln r\rangle + \langle \ln p\rangle\ge$ constant) uncertainty relations for general systems \cite{zozor:pra11,beckner:ams75} and, when known, the improved corresponding relations for central potentials which have been recently found \cite{rudnicki_12,sanchezmoreno:njp06}, what is a further checking of our results.

Second, we find the explicit values of the leading term of the Shannon entropy of Rydberg atoms in both position (see Eq. (\ref{eq:S_rho_res_n})) and momentum (see Eq. (\ref{eq:S_gamma_res_n})) spaces which control the spreading of the charge and momentum distribution of these systems. To do that we have used some sophisticated methods of approximation theory to solve the involved logarithmic functionals of Laguerre and Gegenbauer polynomials. These values satisfy not only the general Shannon-entropy-based uncertainty relation \cite{bialynickibirula:cmp75}, but also the improved corresponding relation for central potentials \cite{rudnicki_12}.

Third, we give the exact values of the position and momentum Fisher informations of Rydberg atoms, which quantify the gradient content of their corresponding wavefunctions. Here, as well, it is obseved that the concomitant Fisher product $I[\rho]\times I[\gamma]$ satisfies the Fisher-information-based uncertainty relation of quantum systems subject to central potentials \cite{romera:cpl05,romera:jmp06}.

Fourth, the complexity measures of Cr\' amer-Rao, Fisher-Shannon and LMC types of hydrogenic Rydberg states, which grasp two-fold global aspects of the corresponding quantum-mechanical density, are  evaluated to first order in both positions and momentum spaces.

Fifth, the previous rigorous findings are applied to not only the $ns$-states, but also to the (experimentally accesible) circular and quasicircular states of Rydberg atoms, what allows one to tackle not only the high anisotropy of the corresponding electron distributions but also the basic classical-quantum correspondence.

Finally, it is worth pointing out that none of the uncertainty relations (generalized Heisenberg-like, logarithmic, entropic)  depends on the nuclear charge $Z$, in accordance with the general observations about entropic relations for homogeneous potentials \cite{sen:jcp06}.

\section*{Appendix}

Determination of the integral $I_{n}$ given by (\ref{eq:rydbergint}).
Making the change of variable $y = -2n + 2 +x$, this integral takes the form
\begin{equation}
\label{eq:partint1}
I_{n} = \int_{-2n+2}^{\infty} dy\,\, (2n -2 +y)^{2n-2} e^{-(2n-2)}e^{-y} y^{2} \ln y^{2}
\end{equation}
 
The development of the Newton binomial and the splitting of the integration interval into the subintervals $[-2n +2, 0]$ and $[0, \infty]$ lead to 
\begin{equation}
\label{eq:partint2}
I_{n} = e^{-(2n-2)}\sum_{k=0}^{2n-2}\binom{2n-2}{k} (2n-2)^{2n-2-k}[J_{0}(k) +J_{1}(n,k)]
\end{equation}
with
\begin{equation}
\label{eq:partint3}
J_{0}(k) = \int_{0}^{\infty} dy\,\, y^{k+2}e^{-y}\ln y^{2} = 2 \Gamma(k +3)\psi(k+3)
\end{equation}
and
\begin{eqnarray}
\label{eq:partint4}
J_{1}(n,k) &=& \int_{-2n+2}^{0} dy\,\, y^{k+2}e^{-y}\ln y^{2} \nonumber \\
 &=& \frac{2^{k+4}}{(k+3)^{2}}(1-n)^{k+3} {}_2 F_{2}\left(\begin{matrix}k+3,& k+3\\ k+4,& k+4&\end{matrix}; 2n-2\right) \nonumber \\
 &   &- \Gamma(k+3) +\Gamma(k+3, -2n+2)\ln (2n-2)^{2}\, ,
\end{eqnarray}
where ${}_2 F_{2}(a_{1}, a_{2}; b_{1}, b_{2}; z)$ is a generalized hypergeometric series \cite{olver_10,sanchezmoreno:njp06}, and $\Gamma(a,z)$ denotes the incomplete gamma function which can be expressed \cite{sanchezmoreno:njp06} as
\begin{equation*}
\Gamma(a,z) = \Gamma(a) - \frac{z^{a}}{a} {}_1 F_{1}(a;a+1;-z), \quad -a\notin \mathbb{N}
\end{equation*}
So that
\begin{equation}
\label{eq:partint5}
\Gamma(k+3, -2n+2) = \Gamma(k+3) -\frac{(2-2n)^{k+3}}{(k+3)} {}_1 F_{1}\left(\begin{matrix}k+3\\ k+4&\end{matrix}; 2n-2\right) 
\end{equation}

Then, the combination of (\ref{eq:partint2})-(\ref{eq:partint5}) allows us to obtain the following expression 

\begin{eqnarray}
\label{eq:partint6}
I_{n}&=& (2n-2)^{2n-2} e^{-(2n-2)}\sum_{k=0}^{2n-2}\binom{2n-2}{k}\frac{1}{(2n-2)^{k}}\times \nonumber\\
	& & \{2\Gamma(k+3)\psi(k+3) + \frac{2^{k+4}(1-n)^{k+3}}{(k+3)^{2}} {}_2 F_{2}\left(\begin{matrix}k+3,& k+3\\ k+4,& k+4&\end{matrix}; 2n-2\right) \nonumber \\
         & & -\frac{(2-2n)^{k+3}}{(k+3)} {}_1 F_{1}\left(\begin{matrix}k+3\\ k+4&\end{matrix}; 2n-2\right)\ln (2n-2)^{2}\},
\end{eqnarray}
where $\psi(z)$ and ${}_1 F_{1}(a; b; z)$ denote the psi or digamma function and the so-called degenerate hypergeometric function, respectively \cite{olver_10, sanchezmoreno:njp06}.

\section*{Acknowledgments}
The authors gratefully acknowledge the Spanish MINECO grant FIS2011-24540 and the excellence grants FQM-1735 and 2445 of the Junta de Andaluc\'ia. They belong to the Andalusian research group FQM-207 (P.S.M, I.V. and J.S.D.) and FQM-239 (S.L.R.). S.L.R. and P.S.M. are grateful for the GENIL-PYR-2010-27 research grant.


\end{document}